\documentclass[showpacs,a4paper,twoside,prb,twocolumn]{revtex4}
\usepackage{graphicx}%
\usepackage{dcolumn}
\usepackage{bm}

\makeatletter
\def\btt#1{\texttt{\@backslashchar#1}}%
\DeclareRobustCommand\bblash{\btt{\@backslashchar}}%
\makeatother

\begin{document}

\preprint{0175-a.tex}
\title{Domain orbital state in ferromagnetic insulating La$_{0.825}$Ca$_{0.175}$Mn$_{0.99}$$^{57}$Fe$_{0.01}$O$_3$ compound}

\author{M. Pissas, G. Papavassiliou, E. Devlin and A. Simopoulos}
\affiliation{Institute of Materials Science, NCSR,  Demokritos,
15310 Aghia Paraskevi, Athens, Greece}
\author{V. Likodimos}
\affiliation{Department of Applied Mathematics and Physics,
National Technical University of Athens, 157 80 Athens, Greece}

\date{\today}
\begin{abstract}
Using M\"ossbauer and EPR spectroscopy we have studied the
insulating ferromagnetic
La$_{1-x}$Ca$_x$Mn$_{0.99}$$^{57}$Fe$_{0.01}$O$_3$ ($x=0.175$)
compound prepared in air and reduced atmosphere. The average
hyperfine field follows a mean field approximation solution in
contrast with the ferromagnetic metallic regime where it
significantly deviates from the order parameter deduced from
neutron diffraction data or mean field approximation. Although the
magnetic measurements show remarkable differences between air
prepared and reduced R samples the corresponding M\"ossbauer
spectra are almost identical. The strong temperature dependance of
the hyperfine field distribution at the $^{57}$Fe nucleus is
related with the supertransferred magnetic field between ferric
ion and the oxygen bridged six nearest-neighbor manganese ions.
The sudden increasing of the width of the hyperfine field
distribution above $T_B\approx 100$ K has been attributed to the
orbital disordering occurring for $T>T_B$.
\end{abstract}
 \maketitle

Apart from the large number of works dedicated in the study of
La$_{1-x}$Ca$_x$MnO$_3$ compound the detailed elucidation of its
ground state is yet an unresolved problem. The two end-compounds
$x=0$ and $x=1$ are A and G-type antiferromagnets respectively. As
Ca substitutes for La in LaMnO$_3$ compound, for $0\leq x\leq
0.125$ the samples display an insulated canted antiferromagnetic
(CA) ground state, changing to  ferromagnetic insulated (FI) for
$0.125\leq x\leq 0.2$ and ferromagnetic metallic (FM) for
$0.23\leq x<0.5$. The FI phase is one of the most unexpected
phases in the physics of manganites since it has not been
predicted neither by the double exchange model nor by any other.
Furthermore, a number of experimental facts show that the ground
state especially in the FI regime is not so simple. In the FI
regime the magnetic measurements reveal, a paramagnetic to
ferromagnetic transition and upon further cooling an additional
anomaly is detected at $T_B\sim 100$ K. Neutron diffraction
measurements\cite{biotteau01} have provided evidences that the
particular anomaly is related with an instability or a
metastability with both magnetic and structural character. In the
case of La$_{1-x}$Sr$_x$MnO$_3$ this transition has been
attributed to an orbital ordered transition leading to a FI
state.\cite{endoh99} $^{57}$Fe M\"ossbauer spectroscopy is an
extremely valuable tool in solid state physics. The long half-life
of the 14.4 keV state yields a resonance linewidth narrow enough
to permit resolution of nuclear fine and hyperfine structure in a
spectrum. Microscopic probes, such as NMR and EPR are often of
limited use,since their signals are  being observable only under
certain conditions, whereas in the case of $^{57}$Fe M\"ossbauer
spectroscopy one is virtually guaranteed of being able to observe
a spectrum.

In manganese perovskites M\"ossbauer spectroscopy in low $^{57}$Fe
and $^{119}$Sn-doped
samples\cite{pissas97,ogale98,tkachuk98,simopoulos98,chechersky99,simopoulos99,kallias99,kallias02,simopoulos01}
has contributed useful information. As an example the sublattice
magnetization can be determined directly from the hyperfine field.
Using M\"ossbauer spectroscopy and electron spin resonance we
investigate the FI regime of the phase diagram with emphasis to
the microscopic origin of the intricate behavior below $T_B$ using
a sample with $x=0.175$ which is in the middle of the
ferromagnetic insulating regime.

\section{Experimental details}
\begin{figure}[htbp]\centering
\includegraphics[angle=0,width=0.7 \columnwidth]{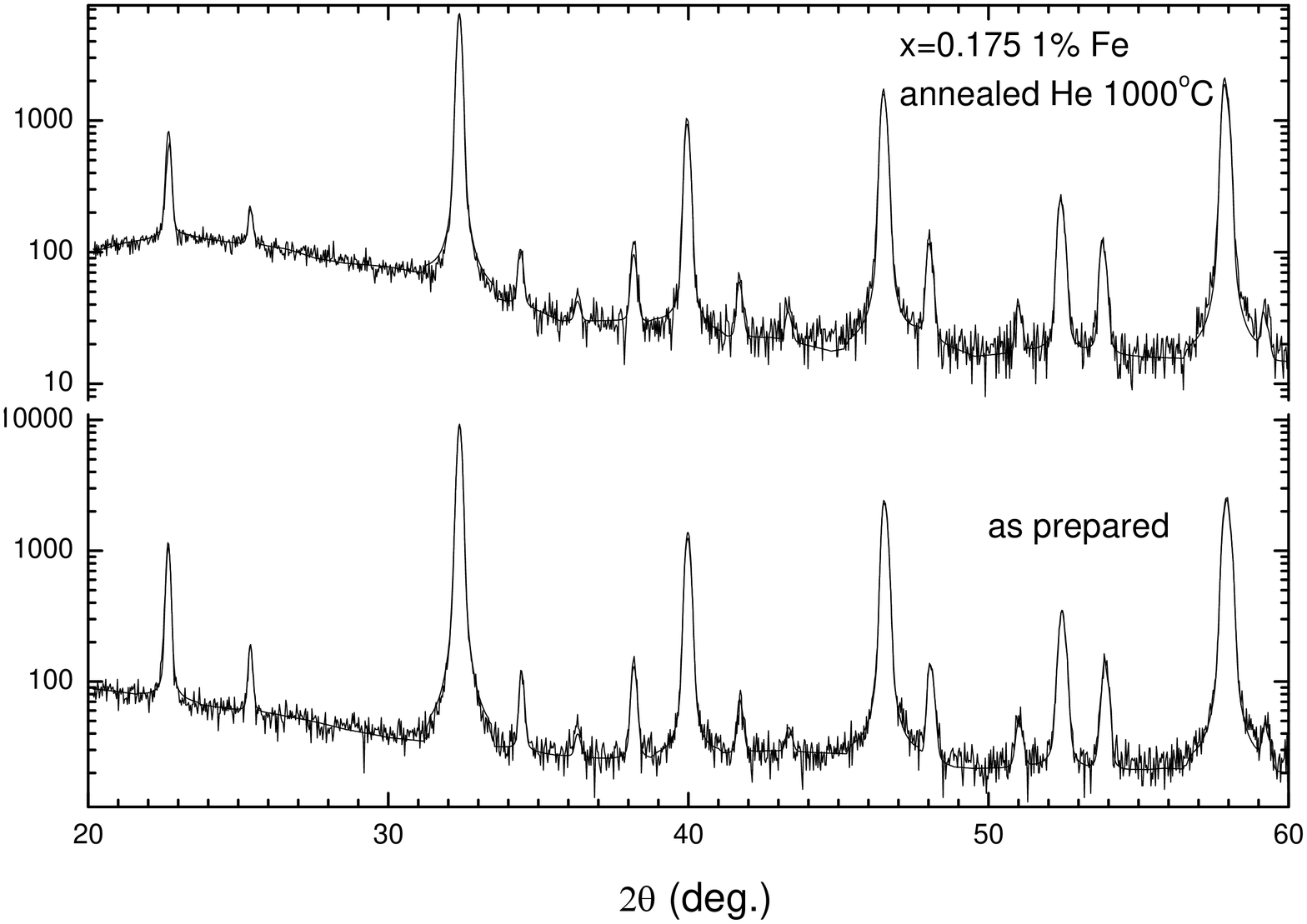}
\caption{X-ray diffraction patterns of (AP) and (R) samples in
semilogarithmic plot.} \label{fig1}
\end{figure}

A sample with nominal composition
La$_{1-x}$\-Ca$_{x}$\-Mn$_{0.99}$\-Fe$_{0.01}$\-O$_3$ ($x=0.175$)
was prepared by the standard solid state reaction method using
Fe$_2$O$_3$ 90\% enriched with $^{57}$Fe. We prepared two samples.
The first sample was prepared in air atmosphere in all stages of
the preparation. We call this sample air prepared sample (AP). The
second sample was annealed in the final stage of the preparation
at 1000$^{\rm o}$ in reduced atmosphere and we call this sample
reduced sample (R). The x-ray diffraction data were analyzed using
the Rietveld refinement method (assuming the orthorhombic $Pnma$
space group for both samples) and revealed single phase materials.
Figure \ref{fig1} shows part of the powder x-ray diffraction
patterns for AP and R samples. The lattice parameters for the AS
and R samples were determined to be $a=5.4883(1)$ \AA ,
$b=7.7585(2)$ \AA , $c=5.5062(1)$ \AA\ and
 $a=5.5012(4)$ \AA, $b=7.7706(4)$ \AA\
and $c=5.5093(4)$ \AA, for AP and R samples, respectively.

The absorption M\"ossbauer spectra (MS) were recorded using a
conventional constant acceleration spectrometer with a
$^{57}$Co(Rh) source moving at room temperature, while the
absorber was kept fixed in a variable temperature cryostat
equipped with a 6.5T superconductive magnet with the field being
perpendicular to the $\gamma-$rays. The resolution was determined
to be $\Gamma /2=0.12$ mm/sec using a thin $\alpha $-Fe foil. DC
magnetization measurements were performed in a SQUID magnetometer
(Quantum Design).

ESR experiments were carried out on a Bruker ER 200D spectrometer
at the X-band (9.41 GHz) with 100 kHz field modulation. The
magnetic field was scaled with a NMR gaussmeter, while temperature
dependent measurements were carried out in the range of 4-300 K
employing an Oxford flow cryostat upon heating from the lowest
temperature. Measurements were performed using either fine
powdered samples dispersed in high vacuum grease or small ceramic
pieces (mass of 1-2 mg), a small portion of which was exposed to
the maximum rf field in order to avoid over-loading the resonant
cavity .\cite{1}
%=================================================================================
%=================================================================================

\begin{figure}[htbp]\centering
\includegraphics[angle=0,width=0.7 \columnwidth]{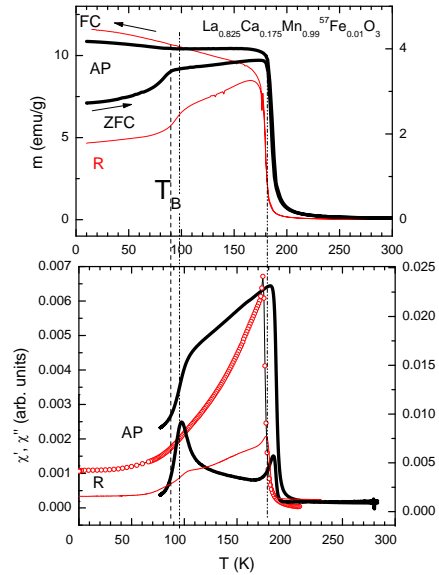}
\caption{(a) Temperature dependence of the magnetic moment at
$H=100$ Oe and (b) and ac susceptibility of AP (thick line) and R
(thin line) samples.} \label{fig2}
\end{figure}

%\begin{figure}[htbp]\centering
%\includegraphics[angle=0,width=\columnwidth]{3.eps}
%\caption{Temperature dependence of the magnetic susceptibility for
%several dc fields of (R) sample.} \label{fig3}
%\end{figure}

%=================================================================================
\section{Magnetic moment measurements}
Figure \ref{fig2}(a) shows the temperature dependence of the
dc-magnetic moment ($m$) in a field of $100$ Oe for the AP and R
samples respectively measured using the SQUID magnetometer. For
both samples the data were collected on heating (zero field
cooling branch ZFC) and on cooling (field cooling branch FC).
Initially the sample was cooled at 4.2 K under zero magnetic
field. Both measurements show a sharp ferromagnetic transition at
180 K and 190 K for the (R) and (AP) sample respectively. In
addition, strong irreversibility between ZFC and FC branch has
been observed for both samples at a temperature ($T_{\rm irr}$),
slightly below $T_c$ Furthermore, the ZFC branch displays a step
like increase with onset temperature $T_B\sim 80$ K and  90 K for
the AP and R samples respectively. Figure \ref{fig2}(b) shows the
corresponding with Figure \ref{fig2}(a) ac-susceptibility
measurements. For both samples the temperature where the
irreversibility starts, in dc measurements, $\chi'$ is reduced,
while the sharp drop which observed in dc measurements is
displayed only by the AP sample. The R sample shows only a
shoulder and only in the $\chi''$ an anomaly is revealed. The
reduction of $\chi'$ for $T<T_{irr}$ sometimes is an indication
for spin glass behavior. However, this behavior can also be
attributed to the domain wall dynamics. Simply the $T_{\rm irr}$
represents the temperature where pining of the domain wall gives
rise to a finite coercivity. Nevertheless, it is difficult to
attribute the sharp drop at $T_B$ to the increasing of the
coercivity. One may argue that the step like increase of the
$m_{\rm ZFC}(T)$ at $T_B$ corresponds to first order transition.
However, the FC branch shows a slope change of the $m_{\rm FC}$ at
$T_B$, indicative for a second order transition. Although we are
not dealing with equilibrium states -as the hysteretic behavior
implies- the overall behavior at $T_B$ is unusual. It is not clear
to the authors why the change of the magnetization in the FC and
ZFC branches is in opposite directions. All the experimental data
advocate for a glassy transition. It is interesting to compare the
present magnetic measurements with those of the
La$_{1-x}$Sr$_x$MnO$_3$ at FI regime occurring at $x=0.125$. This
compound exhibits a cooperative Jahn-Teller first-order transition
at $T_{\rm JT}\approx 270 $ K, first a transition towards a
ferromagnetic and metallic state, at $T_c=181$ K and then, a
magneto-structural first-order transition into a ferromagnetic and
insulating state, at $T_B=159$ K. This FI transition is
characterized by a jump in the
magnetization\cite{endoh99,uhlenbruck99,wagner00,liu01}, typical
for first-order transition, delta function-like variation of the
specific heat, appearance of superstructure peaks, significant
decreasing of the orthorhombicity and the three characteristic
Mn-O distances become very close to each other. In addition Moussa
et al. \cite{moussa03}have found a splitting of the spin waves, an
opening of a gap at ${\bf q}=(0,0,1/2)$ ( $Pnmb$ notation) and a
locking of the spin wave energy on the energy values of phonons.
In a La$_{1-x}$Ca$_x$MnO$_3$ $x=0.175$ sample prepared in reduced
atmosphere we also observed reduction of the othorhombicity of the
O$^/$ phase for $T<T_B$.\cite{pissas04c} This transition was not
observed in the AP sample. All these features occurring at $T_B$,
are indicative for a phase transition, most probably related with
a new orbital/charge order.\cite{endoh99}

\section{M\"ossbauer spectra}
\begin{figure}[htbp]\centering
\includegraphics[angle=0,width=0.6 \columnwidth]{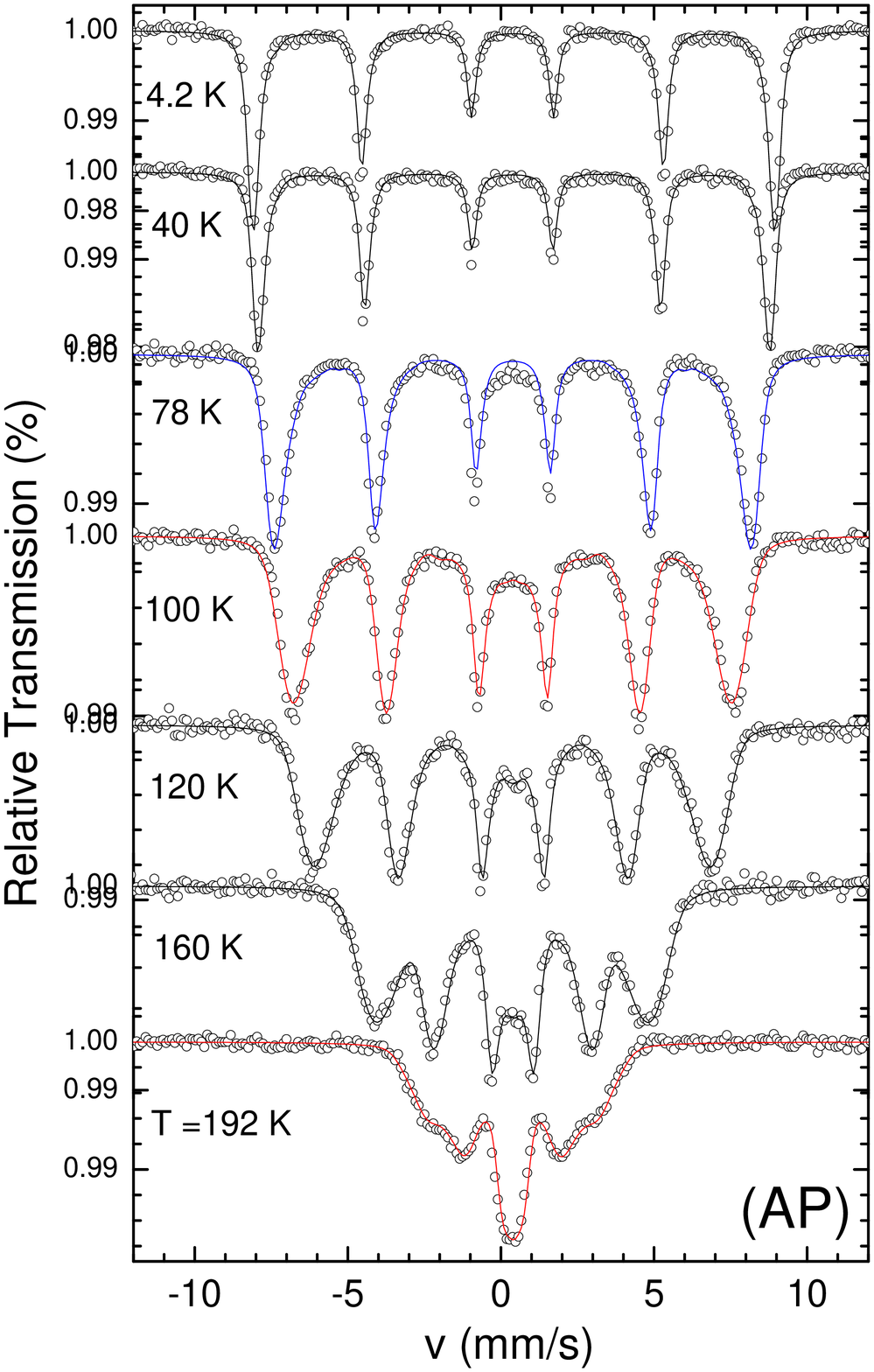}
\caption{M\"{o}ssbauer spectra of (AP) sample.}\label{fig4}
\end{figure}

\begin{figure}[htbp]\centering
\includegraphics[angle=0,width=0.6 \columnwidth]{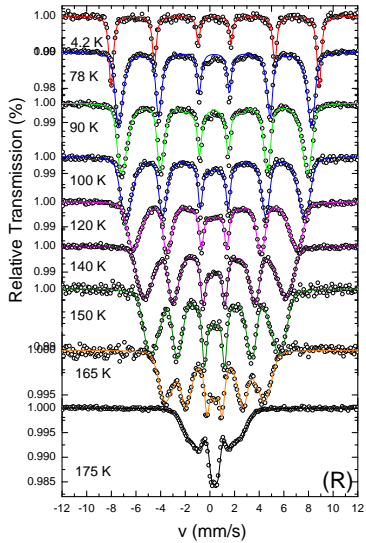}
\caption{M\"{o}ssbauer spectra of (R) sample.} \label{fig5}
\end{figure}

Figures \ref{fig4} and \ref{fig5} show the temperature variation
of the M\"ossbauer spectra of the AP and R samples, respectively.
At $T=300$ K the spectra for both samples consist of a line which
can be fitted by an unresolved doublet, keeping constant the line
width to the value found from the calibration. The isomer shift
$\delta=0.37$ mm/s is characteristic for high spin iron Fe$^{+3}$
in octahedral environment.  The small value of the  quadrupole's
Hamiltonian eigenvalue $\epsilon=0.08$ mm/s indicates that for the
particular $x$ the crystal structure at $T=300$ K does not display
cooperative Jahn-Teller distortion. This value agrees with the
crystallographic data, according to the so-called O$^*$ structure
$(c>a>b/\sqrt{2})$ which is present at $T=300$ K.  In this
structure the octahedra are nearly undistorted and rotated with
respect to the ideal perovskite structure. It is interesting to
note that similar values for $\epsilon$ were found for $x=0.25$
and 0.33 compounds.\cite{pissas97,simopoulos99} As temperature
passes the Curie temperature the spectra become magnetically
split. Near $T_c$ the spectra are rather complicated consisting of
a distribution of hyperfine fields and a paramagnetic component.
By further cooling, the paramagnetic component gradually
disappeared, whereas the width of the hyperfine field distribution
decreased. At $T=4.2$ K, for both R and AP samples, only one
sextet is present with hyperfine parameters ($H_{hf}=526(1)$ kOe,
$\delta=0.506(2)$mm/s, $\epsilon=0.038(2)$ mm/s), and
($H_{hf}=530(1)$ kOe, $\delta=0.506(2)$mm/s, $\epsilon=0.023(1)$
mm/s), respectively. These hyperfine parameters are common for
Fe$^{+3}$ in an octahedral coordination and in the high spin state
$S=5/2$.

\begin{figure}[htbp]\centering
\includegraphics[angle=0,width=0.6 \columnwidth]{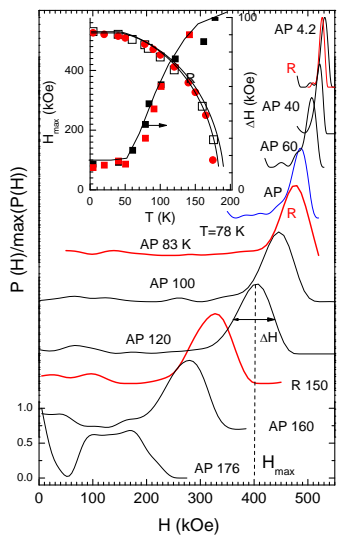}
\includegraphics[angle=0,width=0.6 \columnwidth]{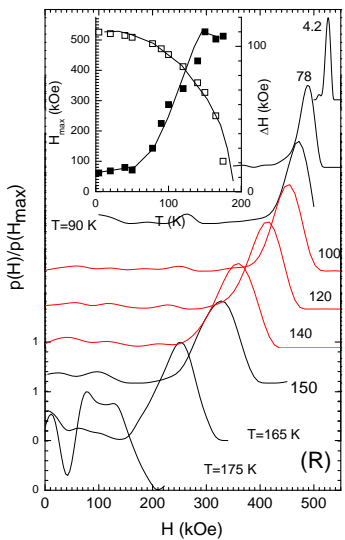}
\caption{Temperature dependence of the magnetic hyperfine field of
for (AP) and (R) samples.} \label{fig6}
\end{figure}

The spectra in the intermediate temperatures were fitted by using
the Le Caer-Dubois  program\cite{caer79}. In this method the
relative transmission $I(i)$ in channel $i (i=1,\cdots N)$ is the
convolution of a continuous hyperfine field distribution, with a
sextuplet of Lorentzian peaks in the thin absorber approximation
theoretically defined by $I(i)=I_0-\int_0^\infty p(H)L(i,H)dH$,
where $I_0$ is the background far from resonance. $L(i,H)$ defines
the contribution in channel $i$ of the elementary sextuplet. After
discretization of the convolution integral
$I(i)=I_0-\sum_{k=1}^{K}\alpha_k\Delta H p(H_k)L(i,H_k)$ the
unknowns $p_k\equiv p(H_k)$ can be determined by the least-squares
minimization of $S=\sum_{i=1}^N W_i (I_e(i)-I(i))^2+\lambda
\int_0^{\infty} (d^2p(H)/dH^2)^2 dH$, subject to $p_k\geq 0$. Here
$I_e(i)$ is the experimental number of counts in channel $i$,
$W_i$ the corresponding weight, and $\lambda$ is the smoothing
parameter. In our case a value 50 has been used for the smoothing
parameter. The resulting hyperfine field distributions $p(H)$ for
both samples are depicted in the insets of Fig. \ref{fig6}. At
$T=4.2$ K $p(H)$ is centered at about 525 kOe with a
FWHM$\equiv\Delta H\approx 15$ kOe. This broadening is present up
to 60 K. For $T>60$ K the FWHM of the $p(H)$ increases linearly
with temperature, up to $T_c$. Interestingly, in order to account
for the region of the spectrum near $v=0$ it is necessary $p(H)$
to be extended down to $H=0$. This part of $p(H)$ is more
pronounced as $T_c$ is approached. Clearly, the temperature
variation of the $\Delta H$ resembled the ZFC branch of the
dc-magnetic moment or the real part of $\chi$. If we accept a
scenario that for $T<T_B$ an orbital ordering occurs, inside the
FI phase, the orbital freedom for $T>T_B$ is responsible for the
temperature dependance of $\Delta H$. At this point we must note
that the temperature variation of $\Delta H$ is different with
that observed in $x=0.33, 0.5$ and 0.6
samples.\cite{kallias99,simopoulos99,kallias02} In these cases
$p(H)$ shows a tail in the low field part. As temperature
increases this tail spreads out to lower fields. In $x=0.175$ case
the main peak broadens but in a "symmetric fashion". The presence
of the component with hyperfine field near zero may arise for
several reasons. Presumably, this part of the distribution
concerns atoms which experience a resultant exchange field via
superexchange interactions with NN spins for which the energy
$g\mu_B H$ is not large with respect to $kT$.  Such atoms belong
to regions where the iron ions destroy the ferromagnetic structure
at a temperature different from this of the undoped sample.
Another explanation can be given supposing that iron behaves as a
magnetic impurity inside a ferromagnetic host, so that the part of
$p(H)$ near $H=0$ represents thermally populates states with
$<H>_T=0$.

%------------------------------------------------------------------
%------------------------------------------------------------------
%Finally, the possibility of superparamagnetic clusters should be
%also considered. In this case, the long range magnetic order
%breaks down to small magnetic clusters which increase in number as
%we approach $T_c$.  Such clusters display paramagnetic peaks with
%$H_{hf}=0$ due to fast relaxation of the magnetic moment of the
%clusters. It should be noted that this behavior is common in all
%manganites doped with Fe (Co) or Sn that have been studied so far
%with M\"{o}ssbauer spectroscopy.(references).It has been also
%observed in the iron containing CMR material FeCr$_2$O$_4$.
%---------------------------------------------------------------------
%----------------------------------------------------------------------

\begin{figure}[htbp]\centering
\includegraphics[angle=0,width=0.6 \columnwidth]{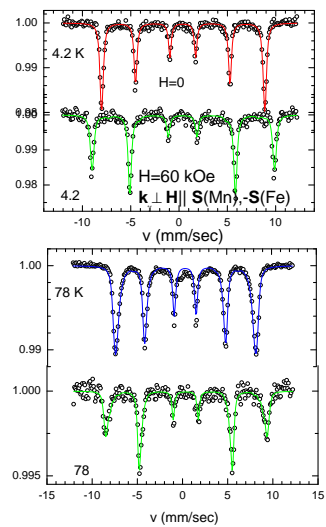}
\caption{M\"ossbauer spectra of R sample in the presence of an
external magnetic field 60 kOe. The external magnetic field is
perpendicular to the $\gamma$-rays. The zero magnetic field
spectra are included in order to permit a direct comparison.}
\label{field}
\end{figure}

We turn now to the microscopic origin of the hyperfine field
distribution as it is revealed in MS. In ferric oxides
contributions to $H_{\rm hf}$ from orbital angular momentum and
conduction-electron polarization are rigorously absent, while
those from dipolar sources are small. This leaves only the contact
field, proportional to the {\bf net} polarization of $s$-electron
density at the nucleus in question, as relevant for these systems.
The contact field ($H_{\rm con}$) is the vector sum of a local
part ${\bf H}_{\rm loc}$ and a supertransferred part ${\bf H}_{\rm
ST}$. ${\bf H}_{\rm loc}$ is proportional to the local $3d$ spin
${\bf S}_0$ on the ion ($S=5/2$ for the case of Fe$^{+3}$) while
${\bf H}_{\rm ST}$ is the resultant contributions from all
single-ligand-bridged  ferric nearest neighbors $n$, each
proportional to the electronic spin ${\bf S}_n$ on the NN cation
site. The resulting field is:
\begin{equation}
{\bf H}_{hf}\approx H_{con}=-C({\bf S}_0/S)+\sum_n B_n({\bf
S}_n/S)
\end{equation}
where $C$ and $B_n$ are positive scalar
parameters.\cite{sawatzky74} The $B_n$ parameters are associated
with the geometry of coordination and can be expressed as a
function of the Fe-O-Fe or Fe-O-Mn bond angle $\phi_n$, namely
$B_n=H_{\pi}+(H_{\sigma}-H_{\pi})\cos^2\phi_n$. In this equation
the fields $H_{\pi,\sigma}$ arise from overlap distortions of the
Fe cation $s$ orbitals caused by the ligand $p$ orbitals having
been unpaired by spin transfer via $\pi$ and $\sigma$ bonds into
unoccupied $3d$ orbitals on the NN cations $n$. In the case of
insulating manganites although the mean value of $H_{\rm hf}$ is
dominated by $H_{\rm loc}$, the fluctuations $\Delta H_{\rm hf}$
which generate the distribution of $H_{\rm hf}$ about its average
are almost exclusively due to fluctuations in the supertransferred
field. In our case $H_{\rm loc}+H_{\rm ST}=-530$ kOe (-the minus
sign means that $H_{\rm hf}$ is antiparallel to the iron spin). By
virtue of theoretical calculations\cite{sawatzky74} it has been
deduced that $H_{\rm loc}\approx -450$ kOe in octahedral oxygen
coordinated ferric iron, a fact implying that the iron moment must
be antiferromagneticaly coupled with the six nearest-neighbor
manganese ions. Most importantly, if the Fe is ferromagnetically
coupled with NN Mn ions, then this coupling will produce positive
supertransferred field resulting in $H_{\rm hf}<450$ kOe, contrary
to the experimental findings.
%-----------------------------------------------------------------
This antiferromagnetic coupling is experimentally verified by
taking spectra in the presence of an external field (vide infra).
%------------------------------------------------------------------
Therefore, the abrupt increase of the width of the hyperfine field
distribution is related to the change in fluctuations in the
supertransferred field. We speculate that these fluctuations are
related with the orbital domains or a new orbital state, formed at
$T_B$. Above $T_B$ the manganese $e_g$ orbitals fluctuate due to
the orbital disorder, producing significant fluctuations in the
supertransferred  field. Oppositely, below $T_B$ orbitaly-ordered
domains are formed leading to freezing of the orbital disorder and
subsequent reduction of the supertransferred magnetic field
fluctuations. Our conclusions are further supported by M\"ossbauer
results on the LaMn$_{0.99}$Fe$_{0.01}$O$_3$, which display the
so-called A-antiferromagnetic structure with four ferromagnetic
and two antiferromagnetic bonds. In this case $H_{\rm
hf}(0)\approx 450$ kOe,\cite{pissas04} due to the positive
contribution of the four (planar) ferromagnetic bonds and the
negative one of the two (apical) antiferromagnetic bonds to the
supertransferred field. It is noticeable that while the magnetic
measurement of the AP sample is different from that of R sample,
the resulting hyperfine distributions are practically similar.
%---------------------------------------------------------
%we need to discuss the above analysis since the Hhyp is 448 kOe and not 470
%for the LMO
%------------------------------------------------------------
At this point we would like to discuss our M\"ossbauer results in
comparison with the NMR ones. NMR spectra at 4.2 K for
La$_{1-x}$Ca$_x$MnO$_3$ ($x\approx 0.175$) show a rich line shape.
Basically, the line shape consists of two peaks corresponding to
Mn$^{+3}$, Mn$^{+4}$ and broad spectral features at intermediate
frequencies attributed to mixed Mn valence states. Our results
clearly show at $T=4.2$ K only one sextet which corresponds to
high spin ($S=5/2$) Fe$^{+3}$ state. In addition, the line width
corresponds to small spread of the hyperfine field in the iron
site. Basically iron sees though the exchange field ${\bf H}_{\rm
ex}=\sum_{n=1}^6 J_n {\bf S}_{n}$ a vector sum of the six NN
manganese spins ($J_n$ is the exchange constant). Broad
distribution of the exchange fields, at $T=4.2$ K, is expected
only if there are several NN Mn$^{3+}$,Mn$^{4+}$ configurations.
As the spectrum at $T=4.2 $ K shows, the fluctuations of the
exchange field are near the resolution of the M\"ossbauer
spectroscopy, a fact implying that the iron sees a unique NN
configuration. Of course we have supposed that the iron has only
manganese ions as NN, an approximation which is good so bad for
1\% iron substituted for Mn ions.
%------------------------
%------------------------
Figure \ref{field} displays spectra taken in the presence of an
external magnetic field of 60 kOe directed perpendicular to the
propagation of $\gamma-$rays at selective temperatures. We notice
that the spectrum at $T=4.2$ K corresponds to a hyperfine field of
588 kOe, which is larger by 60 kOe than the spectrum taken at zero
external field. Furthermore the intensity of the absorption peaks
are in the ratio of 3:4:1 indicating that the Fe magnetic moments
are in the direction of the external magnetic field. The increase
of the hyperfine field by 60 kOe on the other hand shows that the
moments (due to the negative sign of the $H_{\rm hf}$ are
antiparallel to the external field. This can happen only if the Fe
moments are antiferromagnetically coupled to the manganese moments
(which in turn are parallel to the external field). A similar
situation was encountered in the x=0.33 system \cite{simopoulos99}
while for the $x=0.5$ system\cite{kallias99} the Fe moment was
ferromagnetically coupled to the Mn magnetic moments. The
hyperfine field at higher temperatures exceeds the field at zero
external field by more than 60 kOe indicating that the magnetic
order extends far beyond the curie temperature ($T_c\approx 180$
K).
%Finally it is worth noting that the "paramagnetic" peak which
%appears above $\approx~140$ K at zero external field disappears in
%the presence of 60 kOe. This is probably due to the orientation of
%the small spin clusters (superparamagnetic clusters) along the
%external field. %We can %speculate then that the extension of the ordering temperature to
%higher temperatures originates from the organization of small spin
%clusters (magnetic polarons) by the external field.
%-------------

\section{ESR results}
\begin{figure}[htbp]\centering
\includegraphics[angle=0,width=0.6 \columnwidth]{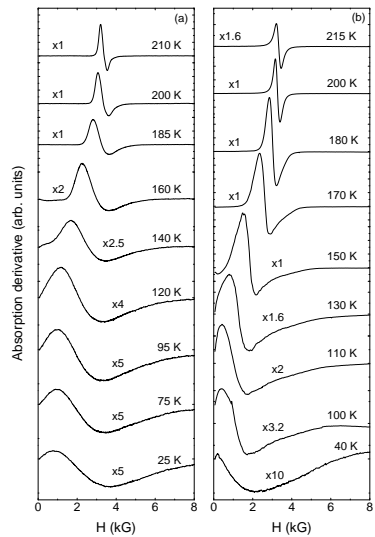}
\caption{ Temperature dependence of the ESR spectra for (a) powder
and (b) bulk samples of the (AP) sample at 9.41 GHz. The scale of
each spectrum is multiplied by the indicated
factors.}\label{7-esr}
\end{figure}

 Figure \ref{7-esr} shows representative ESR spectra of the
AP sample as a function of temperature. A single exchange narrowed
resonance line at $g= 2.0$ due to the strongly coupled
Mn$^{3+}$-Mn$^{4+}$ system \cite{1}, is observed in the
paramagnetic regime. The ESR line broadens and shifts to lower
fields as the FM ordering transition is approached, while a
single, broad ferromagnetic resonance (FMR) mode is observed below
$T_c$, sustained down to the lowest investigated temperature for
both powder and bulk specimens.

\begin{figure}[htbp]\centering
\includegraphics[angle=0,width=0.6 \columnwidth]{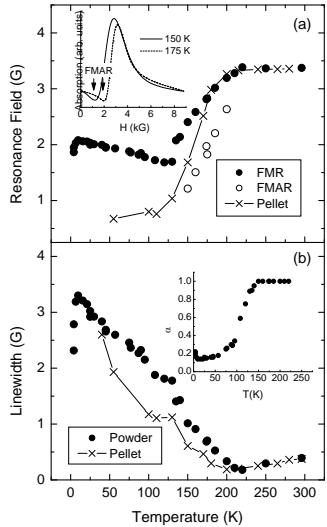}
\caption{Temperature dependence of the (a) resonance field and (b)
linewidth for the (AP)  samples. The inset in (a) shows the dip in
the absorption curve of the powdered samples from which the
antiresonance field (FMAR) is determined. The inset in (b) shows
the temperature variation of the absorption-to-dispersion ratio
(a). } \label{8-esr}
\end{figure}

\begin{figure}[htbp]\centering
\includegraphics[angle=0,width=0.6 \columnwidth]{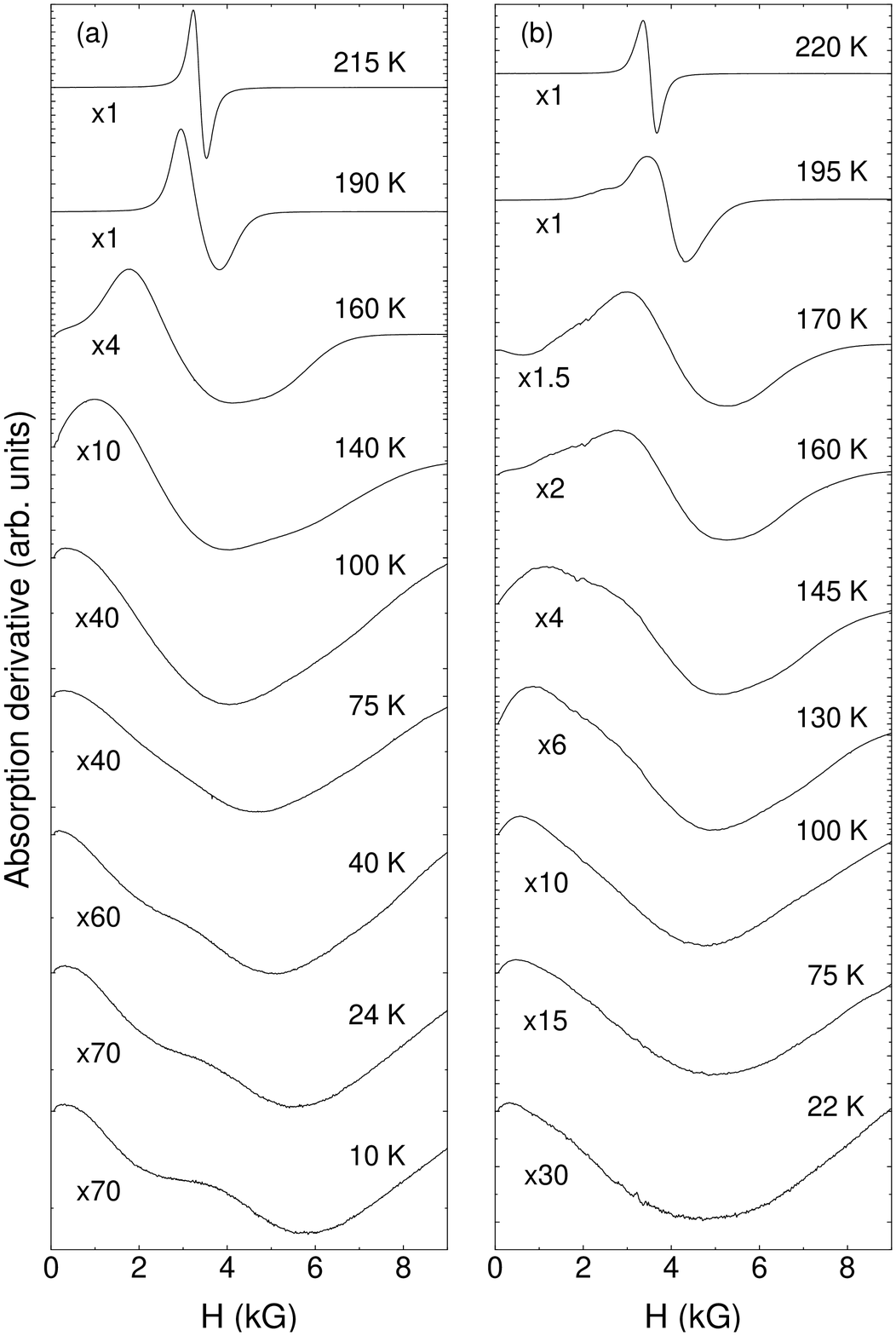}
\caption{Temperature dependence of the ESR spectra for (a) powder
and (b) bulk samples of (R) samples at 9.41 GHz. The scale of each
spectrum is multiplied by the indicated factors. } \label{9-esr}
\end{figure}
\begin{figure}[htbp]\centering
\includegraphics[angle=0,width=0.6 \columnwidth]{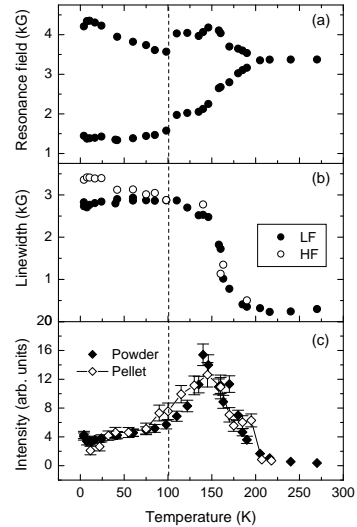}
\caption{Temperature dependence of the (a) resonance fields, (b)
linewidths and (c) the total integrated intensity determined from
the ESR spectra of the powdered (R) samples at 9.41 GHz. }
\label{10-esr}
\end{figure}

The resonance line is well fitted to a single Lorentzian lineshape
in the whole temperature range, including both absorption and
dispersion components to account for the skin effect pertaining
even for powder samples, as well as the tail of the resonance
absorption at negative field, a consequence of the linearly
polarized rf field that becomes important when the width becomes
comparable to the resonance field. \cite{2} Figure \ref{8-esr}
summarizes the temperature dependence of the resonance field $H_r$
and the linewidth $\Delta H$ (half-width at half-height). A small
shift of $H_r$ is derived already from 250 K, in the paramagnetic
phase, pertinent to the presence of demagnetizing fields and the
built up of internal fields due to anisotropic interactions.
\cite{3} At $T<T_c$, a substantial shift of the FMR mode is
observed down to approximately 120 K, where a minimum of $H_r$ is
reached, most clearly evinced for the powder sample that is less
affected by demagnetizing effects. This temperature variation
complies qualitatively with the combined effects of
magnetocrystalline anisotropy and demagnetizing fields, whose
contribution is expected to be most pronounced for the powder and
the bulk specimens, respectively [Fig. \ref{8-esr}(a)].
Integration of the FMR lineshape yields a dip in the absorption
spectra [inset of Fig. \ref{8-esr}(a)], characteristic of the
ferromagnetic antiresonance (FMAR) mode, which has been previously
 observed in other manganites. \cite{4,5,6,7} The variation of the
corresponding resonance field is included in Fig. \ref{8-esr}(a)
in the temperature range of 150-200 K, where $\Delta H$ is
sufficiently narrow for FMAR to be identified, yet not possible to
be reliably analyzed for specimens whose shape is not well-defined
\cite{8}. The linewidth goes through a minimum in the paramagnetic
phase at $T_{\min}=215$ K$= 1.1T_c$ [Fig. \ref{8-esr}(b)],
reflecting the presence of static correlations well above $T_c$,
rather than any critical behavior. \cite{9,10} Below $T_c$, a
continuous increase of $\Delta H$ is observed that saturates for
the powder specimen only at 15 K, suggestive of magnetic
inhomogeneity.\cite{11} Previous studies have shown that a spread
of $T_c$, hardly observable by other methods, may account for the
peak in the FMR linewidth near $T_c$, \cite{12}, while
demagnetizing effects due to pores and surface irregularities in
ceramic materials could be the main source of inhomogeneous
broadening of $\Delta H$ at $T<T_{\min}$.\cite{13} Such a behavior
that is proportional to the average magnetization may accordingly
account for the linewidth broadening down to $T=100$ K, where the
magnetization nearly saturates. However, these effects can not
explain the small, though distinct increase of $H_r$ and the
excessive broadening of the FMR mode, most pronounced for the bulk
sample at $T<100$ K. A substantial increase of the resistivity can
be also inferred at the same temperature from the
absorption-to-dispersion ratio $(\alpha)$ plotted in the inset of
Fig. \ref{8-esr}(b), which decreases rapidly below 150 K and
nearly saturates at 100 K, in agreement with the metal-insulator
transition in this doping range.\cite{14} Different spin dynamics
may be thus invoked at $T<100$ K to explain the FMR temperature
evolution, most likely associated with a disordered FM phase
complying with the marked anomalies of the dc and ac magnetic
measurements. In that case, a different distribution of the
anisotropy axes may also contribute to the FMR linewidth and the
shift of $H_r$, especially for polycrystalline materials. It is
worth noting that a single FMR line has been observed by
high-frequency ESR for low-doped La$_{1-x}$Sr$_x$MnO$_3$ single
crystals at $x=0.15$ and $0.175$, which shifts or splits in two
modes near structural transitions.\cite{7} A single FMR mode has
been also observed for epitaxial LCMO thin films with Mn-excess
and higher Ca content ($x=0.3$), showing an anomalous increase of
$\Delta H$ at low temperatures \cite{16}. On the other hand,
complicated FMR spectra have been reported for powdered LCMO
single crystals with $x=0.18$ in the temperature range of 100-400
K, suggesting the presence of phase separation, that complies
favorably with the FMR data described below for the (R) samples.

Figure \ref{9-esr} shows typical ESR spectra obtained for the (R)
samples at different temperatures. A single ESR line is observed
only at $T>200$ K in the paramagnetic regime, whereas a distorted
lineshape comprising two broad FMR lines at low (LF) and high
field (HF) with respect to $g=2$, is evidenced at lower
temperatures. This two-mode behavior, indicative of increased
magnetic inhomogeneity, was verified for several bulk pieces, an
example being shown in Fig. \ref{9-esr}(b). In this case, a thin
flake of the bulk ceramic (R) sample was measured with the
magnetic field applied perpendicular to its plane, enhancing the
HF line. The FMR spectra for the powder specimen that is less
amenable to demagnetizing effects, could be well fitted using two
Lorentzian lines without appreciable dispersion for $4{\rm
K}<T<200$ K, except for the $T-$range of 150-180 K, where the
lineshape was more distorted [Fig. \ref{9-esr}(a)]. These results
agree qualitatively with the FMR data reported for loose packed
powders of LCMO single crystals with $x=0.18$, where several FMR
lines were resolved down to 150 K, where the conductivity also
attains its maximum value.\cite{14} Figure \ref{10-esr} shows the
corresponding temperature variation of the resonant fields $H_r$,
the linewidth $\Delta H$ and the doubly integrated intensity
including in the latter case that of the bulk piece. The resonance
field of the LF mode, which accounts for about 60-70\% of the
total intensity, varies qualitatively similar to that of the (AP)
sample, reaching an almost constant value below 100 K. Most
importantly, an abrupt shift of both the LF and HF modes occurs at
$T=100$ K towards lower fields [Fig. \ref{10-esr}(a)], accompanied
by the broadening of the total FMR spectrum, mainly resulting from
the HF line [Fig. \ref{10-esr}(b)]. The total intensity reveals a
maximum at $T=145$ K for both specimens, though a shoulder appears
in the vicinity of $T_c$ for the bulk sample, followed by a
relatively small decreasing trend below 100 K [Fig.
\ref{10-esr}(c)], which appears to resemble the ac susceptibility
rather than the dc one, as previously noted.\cite{14} A distorted
FMR lineshape can be, in principle, expected for random powders
with high magnetocrystalline and shape anisotropy.\cite{16}
However, comparison with the FMR and magnetization data of the
(AP) samples clearly points to a dominant thermal treatment
effect.
% Thermogravimetric and structural studies of LCMO have
%shown that the oxygen and the concurrent cation stoichiometry is
%optimized for $x=0.18$ by treatment in air at 1400$^0$C
%,\cite{17}, rendering the {\it (AP) sample more homogeneous and
%less defective than the (R) one}. An enhanced magnetic
%inhomogeneity may be thus inferred for the latter case that
%promotes phase separation, complying with the large thermal
%hysteresis of the dc magnetization and the two-mode FMR spectra.
%Such an explanation could further reconcile similarities with the
%previous FMR results obtained from ground single crystals
%\cite{14} known to be more prone to non-stoichiometry than
%polycrystalline materials.
The persistent anomalies of the FMR spectra for both samples at
$T=100$ K, indicate an intrinsically inhomogeneous magnetic ground
state at low temperatures. This would further correlate with
recent $^{139}$La NMR revealing a freezing transition at $T_f=80$
K \cite{18}, neutron scattering revealing a reentrant structural
transition of the high-temperature pseudo-cubic phase at $T_B=100$
K \cite{biotteau01,pissas04c}, as well as SR experiments showing a
broad maximum of the spin-relaxation at $T=110$ K for Ca doping in
the $x=0.17-0.18$ \cite{20}.

\section{Conclusions}

Stoichiometric samples in the ferromagnetic insulating regime of
La$_{1-x}$Ca$_x$MnO$_3$ compound $0.125\leq x<0.23$ display
characteristic anomalies for $T<100$ K, most probably related with
an orbital transition.  Our M\"{o}ssabuer data revealed that the
La$_{1-x}$Ca$_x$MnO$_3$ ($x=0.175$) sample displays an anomaly in
the temperature variation of the hyperfine field distribution of
the probe $^{57}$Fe nucleus which is related with supertransferred
field. The change in the supertransferred field is closely
connected with a new orbital transition at about $T=100$ K. Our
recent neutron diffraction results revealed\cite{pissas04c} that
only the (R) samples display structural anomaly at $T=100$ K.
Furthermore, our M\"{o}ssbauer spectra and magnetic measurements
show this anomaly in both samples. We attribute this apparent
discrepancy to the fact that this new orbital state in the (AP)
samples is short ranged or glass type. Since the M\"{o}ssbauer
spectroscopy is a local probe it can detect changes that occur in
local level. The conclusion from M\"{o}ssbauer spectra is also in
agreement with EPR data. In the EPR data inhomogeneities are
present in both samples but are most pronounced in the (R) sample
supporting further the interpretation for a new orbital state
which is not long range order in (AP) samples. We believe that the
present results close a gap in the literature concerning the
physics of the (AP) and (R) samples in the ferromagnetic
insulating regime.

\end{document}